\documentclass[preprint,12pt]{elsarticle} 
\usepackage{graphicx} 
\usepackage{amssymb} 
\usepackage{amsmath} 
\usepackage{amsfonts} 
\usepackage{slashed} 
\usepackage[dvipsnames]{xcolor} 
\newcommand{\be}{\begin{equation}} 
\newcommand{\ee}{\end{equation}}

\journal{Nuclear Physics A} 

\begin{document}

\begin{frontmatter}

\title{$\Lambda$ hypernuclear potentials beyond linear density dependence}  

\author[a]{E.~Friedman}\corref{cor1} 
\author[a]{A.~Gal} 
\address[a]{Racah Institute of Physics, The Hebrew University, 9190400 
Jerusalem, Israel} 
\cortext[cor1]{corresponding author: Eli Friedman, 
eliahu.friedman@mail.huji.ac.il}  

\begin{abstract} 

In a recent paper [PLB 837 (2023) 137669] we showed that all measured 
($1s_\Lambda$, $1p_\Lambda$) pairs of $\Lambda$ binding energies in 
$\Lambda$-hypernuclei across the periodic table, $12\leq A \leq 208$, 
can be obtained from a $\Lambda$-nucleus optical potential with only 
two adjustable $\Lambda N$ and $\Lambda NN$ parameters, associated with 
leading linear and quadratic terms in the nuclear density, derived by 
fitting  $^{16}_{~\Lambda}$N binding energies. Here we extend the previous 
analysis by performing least-squares fits to the full set of data points. 
Consequences of suppressing $\Lambda NN$ interactions between `core' nucleons 
and `excess' neutrons are studied and related predictions are made for 
($1s_\Lambda$, $1p_\Lambda$) binding energies in $^{40,48}_{~~~~\Lambda}$K, 
obtainable from upcoming $^{40,48}$Ca($e,e'K^+$) JLab experiments. We find 
$\Lambda$-nucleus partial potential depths of $D^{(2)}_\Lambda = -38.6\pm 
0.8$~MeV ($\Lambda N$) and $D^{(3)}_\Lambda = 11.3\pm 1.4$~MeV ($\Lambda NN$), 
with a total depth $D_\Lambda = -27.3\pm 0.6$~MeV at nuclear-matter density 
$\rho_0$=0.17~fm$^{-3}$, consistently with our previous results. Extrapolation 
to higher nuclear densities and possible relevance to the `hyperon puzzle' in 
neutron-star matter are discussed. 

\end{abstract} 

\begin{keyword}
Hyperon strong interaction. $\Lambda$ hypernuclei. Optical model fits.  
\end{keyword} 

\end{frontmatter}

\section{Introduction}
\label{sec:intro}

$\Lambda$ single-particle states down to the $1s_{\Lambda}$ state established 
in old emulsion studies and in more recent ($K^-,\pi^-$), ($\pi^+,K^+$) and 
($e,e'K^+$) reactions across the periodic table up to $^{208}_{~~\Lambda}$Pb 
provide good evidence for a $\Lambda$-nucleus attractive potential depth 
$D_{\Lambda} \approx -30$~MeV~\cite{GHM16}. Skyrme-Hartree-Fock (SHF) 
studies in terms of a density dependent (DD) $\Lambda$-nuclear potential 
$V_{\Lambda}^{\rm SHF}(\rho)$~\cite{MDG88} concluded that a $\rho^2$ term 
motivated by three-body $\Lambda NN$ interactions provides a large repulsive 
(positive) contribution to the $\Lambda$-nuclear potential depth $D_{\Lambda}$ 
at nuclear-matter density $\rho_0$, $D_{\Lambda}^{(3)}\approx 30$~MeV. 
This repulsive component of $D_{\Lambda}$ is more than  compensated at 
$\rho_0$ by a roughly twice larger attractive depth value, $D_{\Lambda}^{(2)} 
\approx -60$~MeV, induced by a two-body $\Lambda N$ interaction. Both values 
of $D_{\Lambda}^{(2)}$ and $D_{\Lambda}^{(3)}$ are excessive, as discussed 
below. We note that $D_{\Lambda}=D_{\Lambda}^{(2)}+D_{\Lambda}^{(3)}$ is 
defined as $V_{\Lambda}(\rho_0)$ in the limit $A\to\infty$ at a given 
nuclear-matter density $\rho_0$, with a value 0.17~fm$^{-3}$ assumed here. 
As nuclear density is increased beyond nuclear-matter density $\rho_0$, the 
balance between attractive $D_\Lambda^{(2)}$ and repulsive $D_\Lambda^{(3)}$ 
tilts towards the latter. This may result in nearly total expulsion of 
$\Lambda$ hyperons from neutron-star matter at density several times $\rho_0$, 
suggesting a nucleonic equation of state sufficiently stiff to support two 
solar-mass neutron stars, thereby providing one possible solution to the 
`hyperon puzzle'~\cite{Vidana22}. However, there is no guarantee that 
three-body $\Lambda NN$ interactions are universally repulsive. 

Here, as well as in our preceding work~\cite{FGa23,FGa22}, we construct 
a physically appropriate DD $\Lambda$-nucleus optical potential 
$V_{\Lambda}^{\rm OPT}(\rho)=V_{\Lambda}^{(2)}(\rho)+V_{\Lambda}^{(3)}(\rho)$, 
where $V_{\Lambda}^{(2)}$ ($V_{\Lambda}^{(3)}$) is associated with two-body 
(three-body) $\Lambda N$ ($\Lambda NN$) interactions. Our aim is to establish 
the relative importance of the two parts of the potential by employing as 
simple model as possible, with extrapolation to higher densities in mind. We 
follow the DD optical potential approach applied by Dover-H\"{u}fner-Lemmer 
to pions in nuclear matter \cite{DHL71}. For the $\Lambda$-nucleus system, 
it provides expansion in powers of the nuclear density $\rho(r)$, consisting 
of two components: (i) $V_{\Lambda}^{(2)}(\rho)$, a two-body $\Lambda N$ 
interaction linear-density term modified by a long-range Pauli correlation 
factor that superposes powers of the Fermi momentum $k_F \propto \rho^{1/3}$ 
starting at $\rho^{4/3}$, and (ii) $V_{\Lambda}^{(3)}$, a short-range $NN$ 
correlation factor dominated in the present context by a three-body $\Lambda N
N$ interaction term starting at $\rho^2$. These two components were determined 
in our preceding work by fitting to $\Lambda$ binding 
energies ($B_{\Lambda}$) of the $1s_{\Lambda}$ and $1p_{\Lambda}$ states in 
$^{16}_{~\Lambda}$N and then tested by the extent to which 
such a fit reproduces all other $B^{1s,1p}_{\Lambda}$ values in heavier 
$\Lambda$ hypernuclei across the periodic table up to $^{208}_{~~\Lambda}$Pb. 
The introduction of Pauli correlations, a must in any $G$-matrix version 
of optical-potential calculations, reduces considerably the size of $D_{
\Lambda}^{(2)}$ reported in SHF calculations, as seen in Table~\ref{tab:SHF}. 
It affects also higher powers of $\rho$, in particular $\rho^2$ contributions, 
thereby leading to a sizable reduction of $D_{\Lambda}^{(3)}$ from any of the 
No-Pauli values listed in the table to the Yes-Pauli value listed in the last 
line.  

\begin{table}[!htb]
\begin{center}
\caption{$\Lambda$-nuclear potential depths (in MeV) from two SHF calculations 
fitting $B_{\Lambda}$ data points and from our own optical-model two-parameter  
fit to $B_{\Lambda}^{1s,1p}(^{16}_{~\Lambda}$N) values. Pauli-Yes (Pauli-No) 
stands for including (excluding) nuclear Pauli correlations in 
$V_{\Lambda}^{(2)}(\rho)$.}
\begin{tabular}{cccccc}
\hline
Method & Pauli & Data Points & $D_{\Lambda}^{(2)}$ &
$D_{\Lambda}^{(3)}$ & $D_{\Lambda}$ \\
\hline
$V_{\Lambda}^{\rm SHF}$~\cite{MDG88} & No  &  3 & $-$57.8 & 31.4 & $-$26.4  \\
$V_{\Lambda}^{\rm SHF}$~\cite{SHi14} & No  & 35 & $-$55.4 & 20.4 & $-$35.0  \\
$V_{\Lambda}^{\rm OPT}$~\cite{FGa23} & No  &  2 & $-$57.6 & 30.2 & $-$27.4  \\
$V_{\Lambda}^{\rm OPT}$~\cite{FGa23} & Yes &  2 & $-$41.6 & 13.9 & $-$27.7  \\
\hline
\end{tabular}
\label{tab:SHF}
\end{center}
\end{table}

\begin{figure}[!htb]
\begin{center}
\includegraphics[width=0.8\textwidth]{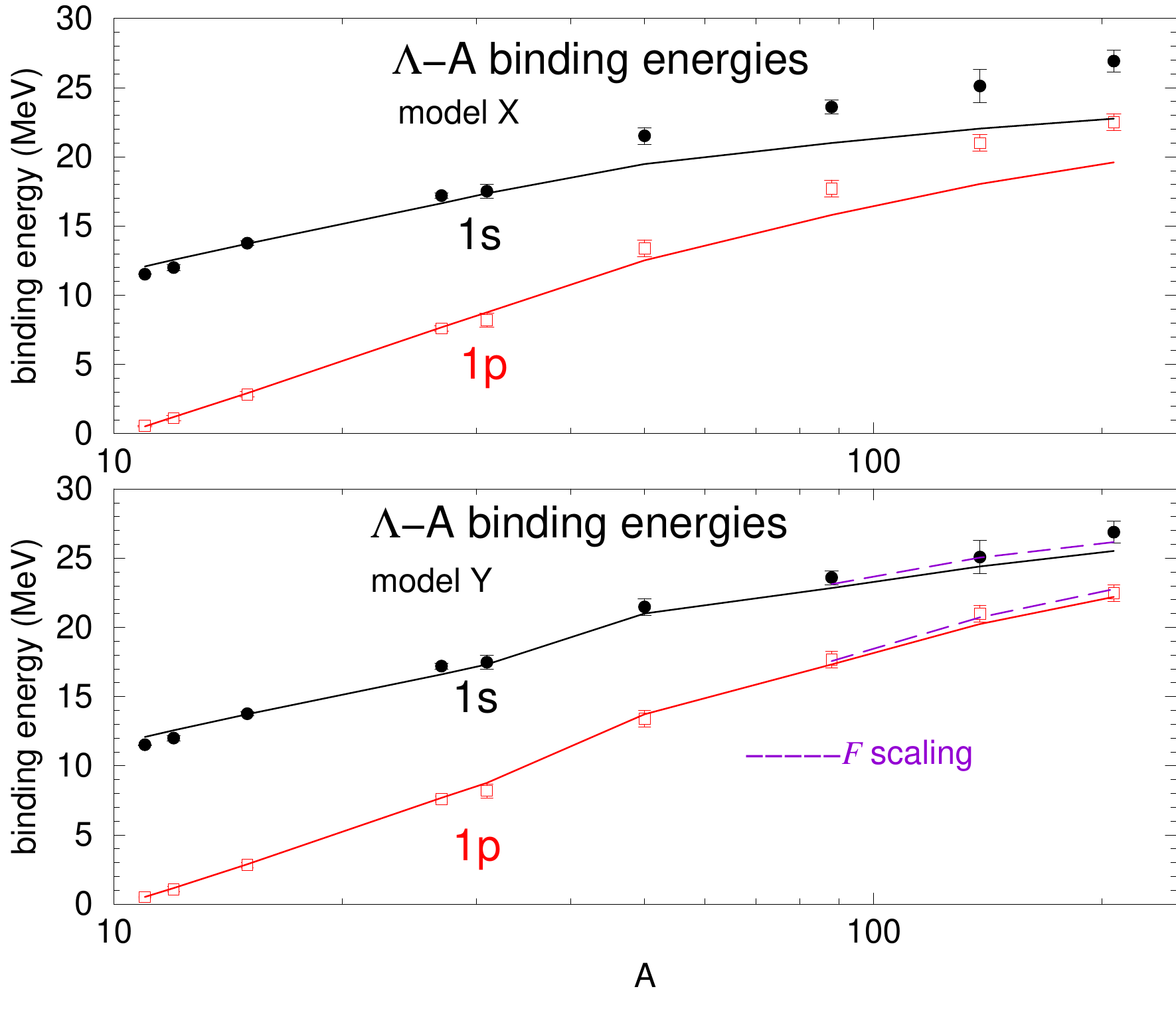}
\caption{$B_{\Lambda}^{1s,1p}(A)$ values across the periodic table as 
calculated in Models X (upper) and Y (lower), compared with data points, 
including uncertainties. Continuous lines connect calculated values. 
Figure updating Fig.~3 in Ref.~\cite{FGa23}. The upper part, Model X, 
uses the full $\rho^2$ term. The lower part, Model Y, same as Model Y$_0$ 
in Fig.~4 (upper) of Ref.~\cite{FGa22}, replaces $\rho^2$ by a reduced form, 
decoupling $N>Z$ excess neutrons from $N=Z$ symmetric-core nucleons, see text. 
The dashed line is for $\rho^2$ replaced by $F\rho^2$, with a suppression 
factor $F$ given by Eq.~(\ref{eq:F}) below.} 
\label{fig:FGa23} 
\end{center} 
\end{figure}

Another important feature established by fitting $B_{\Lambda}$ values across 
the periodic table in terms of $V_{\Lambda}^{\rm OPT}$~\cite{FGa23} is the 
need to suppress the operation of the 3-body $\Lambda N_1 N_2$ $\rho^2$ 
term between nucleons $N_1$ from the $N=Z$ symmetric `core' and neutrons 
$N_2$ from the $N > Z$ neutron `excess'. The microscopic origin of such 
suppression is briefly discussed in Sect.~\ref{sec:method} while its actual 
effect is demonstrated here in Fig.~\ref{fig:FGa23}, by going 
from Model X in the upper part to Model Y in the lower part. Model X uses the 
full $\rho^2$ term of $V_{\Lambda}^{\rm OPT}$ whereas Model Y uses a 3-body 
term appropriately reduced. Model X is seen to lead to substantial 
underbinding of $1s_{\Lambda}$ and $1p_{\Lambda}$ states in $N > Z$ 
hypernuclei. 

In the present work we apply standard least-squares fits to the 
fullbody of $B_{\Lambda}^{1s,1p}(A)$ data considered 
previously by us~\cite{FGa23} rather than fit exclusively 
to the two $B_{\Lambda}^{1s,1p}(^{16}_{~\Lambda}$N) values. We study 
in particular correlations between the two terms of the potential. 
Suppression of $\Lambda NN$ interactions of `core' nucleons with `excess' 
neutrons is shown to be an essential ingredient of optical-potential fits. 
The partial depths of $\Lambda $-nucleus potentials at nuclear density 
$\rho _0$=0.17~fm$^{-3}$ for symmetric nuclear matter and $A\to\infty$ 
are found to be highly correlated, $D^{(2)}_\Lambda=-38.6\pm$0.8~MeV and 
$D^{(3)}_\Lambda$=11.3$\pm$1.4~MeV, with a total potential depth 
$D_\Lambda=-27.3\pm$0.6~MeV. 

The paper is organized as follows. In Sect.~\ref{sec:method} the form of 
the optical potential $V_{\Lambda}^{\rm OPT}(\rho)$ used to evaluate binding 
energy values ($B_{\Lambda}$) is reviewed, followed by a discussion of the 
nuclear densities and $B_{\Lambda}$ data used, both of which are basic input 
to the optical model methodology and its applications. Least-squares fit 
results are given in Sect.~\ref{sec:results} where effects due to the `excess' 
neutrons are studied in detail, and related predictions for the yet unobserved 
$1s_{\Lambda}$ and $1p_{\Lambda}$ states in $^{48,40}_{~~~~\Lambda}$K are 
reported. The validity of the model for higher excited states in heavy nuclei 
is explored, in addition to discussing also light nuclei. Concluding 
remarks are presented in Sect.~\ref{sec:disc}.

\section{Methodology}
\label{sec:method}

\subsection{Optical potential}
\label{subsec:opt}

The optical potential employed in this work, 
$V_{\Lambda}^{\rm OPT}(\rho)=V_{\Lambda}^{(2)}(\rho)+V_{\Lambda}^{(3)}(\rho)$, 
consists of terms representing two-body $\Lambda N$ and three-body $\Lambda 
NN$ interactions, respectively: 
\begin{equation} 
V_{\Lambda}^{(2)}(\rho) = -\frac{4\pi}{2\mu_{\Lambda}}f^{(2)}_A\,
C_{\rm Pauli}(\rho)\,b_0\rho, 
\label{eq:V2} 
\end{equation}  
\begin{equation} 
V_{\Lambda}^{(3)}(\rho) = +\frac{4\pi}{2\mu_{\Lambda}}f^{(3)}_A\,B_0\,
\frac{\rho^2}{\rho_0}, 
\label{eq:V3} 
\end{equation} 
with $b_0$ and $B_0$ strength parameters in units of fm ($\hbar=c=1$). 
In these expressions, $A$ is the mass number of the {\it nuclear core} 
of the hypernucleus, $\rho$ 
is a nuclear density normalized to $A$, $\rho_0=0.17$~fm$^{-3}$ stands for 
nuclear-matter density, $\mu_{\Lambda}$ is the $\Lambda$-nucleus reduced mass 
and $f^{(2,3)}_A$ are kinematical factors transforming $b_0$ and $B_0$ 
from the $\Lambda N$ and $\Lambda NN$ c.m. systems, respectively, to the 
$\Lambda$-nucleus c.m. system: 
\begin{equation} 
f^{(2)}_A=1+\frac{A-1}{A}\frac{\mu_{\Lambda}}{m_N},\,\,\,\,\,\,
f^{(3)}_A=1+\frac{A-2}{A}\frac{\mu_{\Lambda}}{2m_N}. 
\label{eq:fA} 
\end{equation} 
\begin{equation} 
C_{\rm Pauli}(\rho)=(1+\alpha_P\frac{3k_F}{2\pi}f^{(2)}_Ab_0)^{-1}. 
\label{eq:Cpauli} 
\end{equation}
The form of $f^{(2)}_A$ coincides with the way it is used for $V^{(2)}$ in 
atomic/nuclear hadron-nucleus bound-state problems~\cite{FGa07}. Next in 
Eq.~(\ref{eq:V2}) is the DD Pauli correlation function $C_{\rm Pauli}(\rho)$, 
with Fermi momentum $k_F=(3{\pi^2}\rho/2)^{1/3}$. The parameter $\alpha_P$ 
in Eq.~(\ref{eq:Cpauli}) switches off ($\alpha_P$=0) or on ($\alpha_P$=1) 
Pauli correlations in a form suggested in Ref.~\cite{WRW97} and practised 
in $K^-$ atoms studies~\cite{FGa17}. As shown in Ref.~\cite{FGa23}, including 
$C_{\rm Pauli}(\rho)$ in $V_{\Lambda}^{(2)}$ affects strongly the balance 
between the derived potential depths $D_\Lambda^{(2)}$ and $D_\Lambda^{(3)}$. 
However, introducing it also in $V_{\Lambda}^{(3)}$ is found to make little 
difference, which is why it is skipped in Eq.~(\ref{eq:V3}). Finally we note 
that the low-density limit of $V_{\Lambda}^{\rm OPT}$ requires according to 
Ref.~\cite{DHL71} that $b_0$ is identified with the c.m. $\Lambda N$ 
spin-averaged scattering length (positive here). 

\subsection{Nuclear densities}
\label{subsec:rho}

In optical model applications similar to the
one adopted here, it is crucial to ensure that the radial extent of
the densities, e.g., their r.m.s. radii, follow closely values derived
from experiment. With $\rho(r) = \rho_p(r) + \rho_n(r)$, the sum of
proton and neutron density distributions, respectively, we relate
the proton densities to the corresponding charge densities where
the finite size of the proton charge and recoil effects are included.
This approach is equivalent to assigning some finite range to the
$\Lambda$-nucleon interaction. For the lightest elements in our database we
used harmonic-oscillator type densities, assuming the same radial parameters
also for the corresponding neutron densities \cite{Elton61}. For species
beyond the nuclear $1p$ shell we used two-parameter and three-parameter
Fermi distributions normalized to $Z$ for protons and $N=A-Z$ for neutrons,
derived from nuclear charge distributions assembled in Ref.~\cite{AM13}.
For medium-weight and heavy nuclei, the r.m.s. radii of neutron density
distributions assume larger values than those for proton density
distributions, as practiced in analyses of exotic atoms \cite{FGa07}.
Furthermore, once neutrons occupy single-nucleon orbits beyond those occupied
by protons, it is useful to represent the nuclear density $\rho(r)$ as
\begin{equation} 
\rho(r)=\rho_{\rm core}(r)+\rho_{\rm excess}(r), 
\label{eq:exc1} 
\end{equation}
where $\rho_{\rm core}$ refers to the $Z$ protons plus the charge symmetric
$Z$ neutrons occupying the same nuclear `core' orbits, and $\rho_{\rm excess}$
refers to the $(N-Z)$ `excess' neutrons associated with the nuclear periphery. 

One of the conclusions of Ref.~\cite{FGa23} was that a straightforward 
application of Eqs.~(\ref{eq:V2}) and (\ref{eq:V3}) across the periodic table, 
with $b_0$ and $B_0$ fitted to $B_{\Lambda}^{1s,1p}$ values in light 
hypernuclei, leads to substantially underbound $1s_\Lambda$ and $1p_\Lambda$ 
states calculated in heavier hypernuclei, as shown in the upper part of 
Fig.~\ref{fig:FGa23} above. This underbinding was avoided by discarding the 
bilinear term $2\rho_{\rm core}\rho_{\rm excess}$ in $\rho ^2$ when excess 
neutrons occupy shell-model orbits higher than those occupied by protons, 
as shown in the lower part of Fig.~\ref{fig:FGa23}. The microscopic origin 
of the suppression ansatz applied here is traced back to $\Lambda NN$ 
pion-exchange models that couple the isospin $T=0$ 
$\Lambda$ hyperon to the $T=1$ $\Sigma$ and $\Sigma^{\ast}(1385)$ 
hyperons~\cite{Spi58,BLN67}, as suggested also in modern $\chi$EFT 
models~\cite{Petsch16} and practised in Ref.~\cite{GKW20}. 
The resulting $\Lambda NN$ potential contributions depend then on the nucleon 
isospins through a ${\vec\tau}_1\cdot{\vec\tau}_2$ factor which vanishes in 
direct matrix elements when $N_1$ runs over $T=0$ closed-shell core nucleons 
and $N_2$ is an excess neutron. The exchange partners of such matrix elements 
renormalize the two-body $\Lambda N$ interaction~\cite{GSD71}.

In the spirit of Ref.~\cite{FGa23} and of the present work, namely, 
avoiding explicit models as much as possible, we replace $\rho ^2$ by 
$\rho_{\rm core}^2+\rho_{\rm excess}^2$, represented by
\begin{equation}
\rho_{\rm core}^2+\rho_{\rm excess}^2\rightarrow(2\rho_p)^2+(\rho_n-\rho_p)^2,
\label{eq:exc2}
\end{equation}
in terms of the available densities $\rho_p$ and $\rho_n$. 
It is straightforward to show that the volume integral of 
$(2\rho_p)^2+(\rho_n-\rho_p)^2$ is equal to $F$ times the
volume integral of $\rho ^2$ where
\begin{equation}
F=\frac{(2Z)^2+(N-Z)^2}{A^2}.
\label{eq:F}
\end{equation}
Using $F\rho ^2$ in $V_{\Lambda}^{(3)}(\rho)$ to suppress the bilinear term, 
instead of using Eq.~(\ref{eq:exc2}), leads to almost the same calculated 
binding energies, as also shown in the lower part of Fig.~\ref{fig:FGa23}.

\subsection{$\Lambda$ binding energy data}
\label{subsec:B_L}

For a $B_{\Lambda}$ data base we used {\it all} experimentally available 
($1s_{\Lambda},1p_{\Lambda}$) pairs of single-$\Lambda$ states in 
$\Lambda$~hypernuclei. Obviously the quoted uncertainty of the energy is 
a major factor, but due to the availability of experimental results from 
different sources the consistency of various data had also to be considered. 
The $B_{\Lambda}$ data base chosen in Ref.~\cite{FGa23} and employed also 
in the present work is based on Table~IV of Ref.~\cite{GHM16} and is given 
in Table~\ref{tab:data} here, listing also the corresponding strangeness 
production reactions in which single-$\Lambda$ states were identified. 
Most of these states were derived from ($\pi^+,K^+$) spectra, where nuclear 
excitation admixtures often affect the extracted $B_{\Lambda}$ values. Some 
of the implied systematical uncertainties are discussed in Ref.~\cite{GHM16}, 
and for $p$-shell $\Lambda$ hypernuclei also in Ref.~\cite{BBF17}. 
Note that $1p_{\Lambda}$ binding starts at $A=12$, so $^{12}_{~\Lambda}$B 
is the lightest hypernucleus considered in our fits. Its extremely small 
$\delta B_{\Lambda}$ uncertainty values were increased by us to $\pm$0.2 MeV, 
making the $B_{\Lambda}^{1s,1p}(^{12}_{~\Lambda}$B) values consistent with 
their corresponding values in the charge-symmetric $^{12}_{~\Lambda}$C 
hypernucleus. Generally in this mass range, charge symmetry breaking incurs 
an uncertainty of $\approx\,$0.1~MeV~\cite{Gal23}. 

\begin{table}[!htb]
\begin{center}
\caption{$1s_\Lambda$ and $1p_\Lambda$ binding energies (MeV) in hypernuclei 
$_\Lambda^A$Z, including uncertainties, deduced from several strangeness 
production reactions (SPR) as listed in Table IV of Ref.~\cite{GHM16}. 
These are the $B_\Lambda$ values considered in our preceding 
work~\cite{FGa23} as well as in the present work.} 
\label{tab:data} 
\centering
\begin{tabular}{ccllll} 
\hline
$_\Lambda^A$Z & SPR & ~~$B^{1s}_\Lambda$ & $~~\pm$ & $B^{1p}_\Lambda $ & 
$~~\pm $ \\ 
\hline
$^{12}_{~\Lambda}$B    & ($e,e'K^+$)   & 11.52 & 0.02 & 0.54  & 0.04  \\ 
$^{13}_{~\Lambda}$C    & ($\pi^+,K^+$) & 12.0  & 0.2  & 1.1   & 0.2   \\
$^{16}_{~\Lambda}$N    & ($e,e'K^+$)   & 13.76 & 0.16 & 2.84  & 0.18  \\ 
$^{28}_{~\Lambda}$Si   & ($\pi^+,K^+$) & 17.2  & 0.2  & 7.6   & 0.2   \\
$^{32}_{~\Lambda}$S    & ($K^-,\pi^-$) & 17.5  & 0.5  & 8.2   & 0.5   \\
$^{51}_{~\Lambda}$V    & ($\pi^+,K^+$) & 21.5  & 0.6  & 13.4  & 0.6   \\
$^{89}_{~\Lambda}$Y    & ($\pi^+,K^+$) & 23.6  & 0.5  & 17.7  & 0.6   \\
$^{139}_{~~\Lambda}$La & ($\pi^+,K^+$) & 25.1  & 1.2  & 21.0  & 0.6   \\
$^{208}_{~~\Lambda}$Pb & ($\pi^+,K^+$) & 26.9  & 0.8  & 22.5  & 0.6   \\  
\hline
\end{tabular}
\end{center}
\end{table}

\begin{figure}[!h]
\centering
\includegraphics[width=0.8\textwidth]{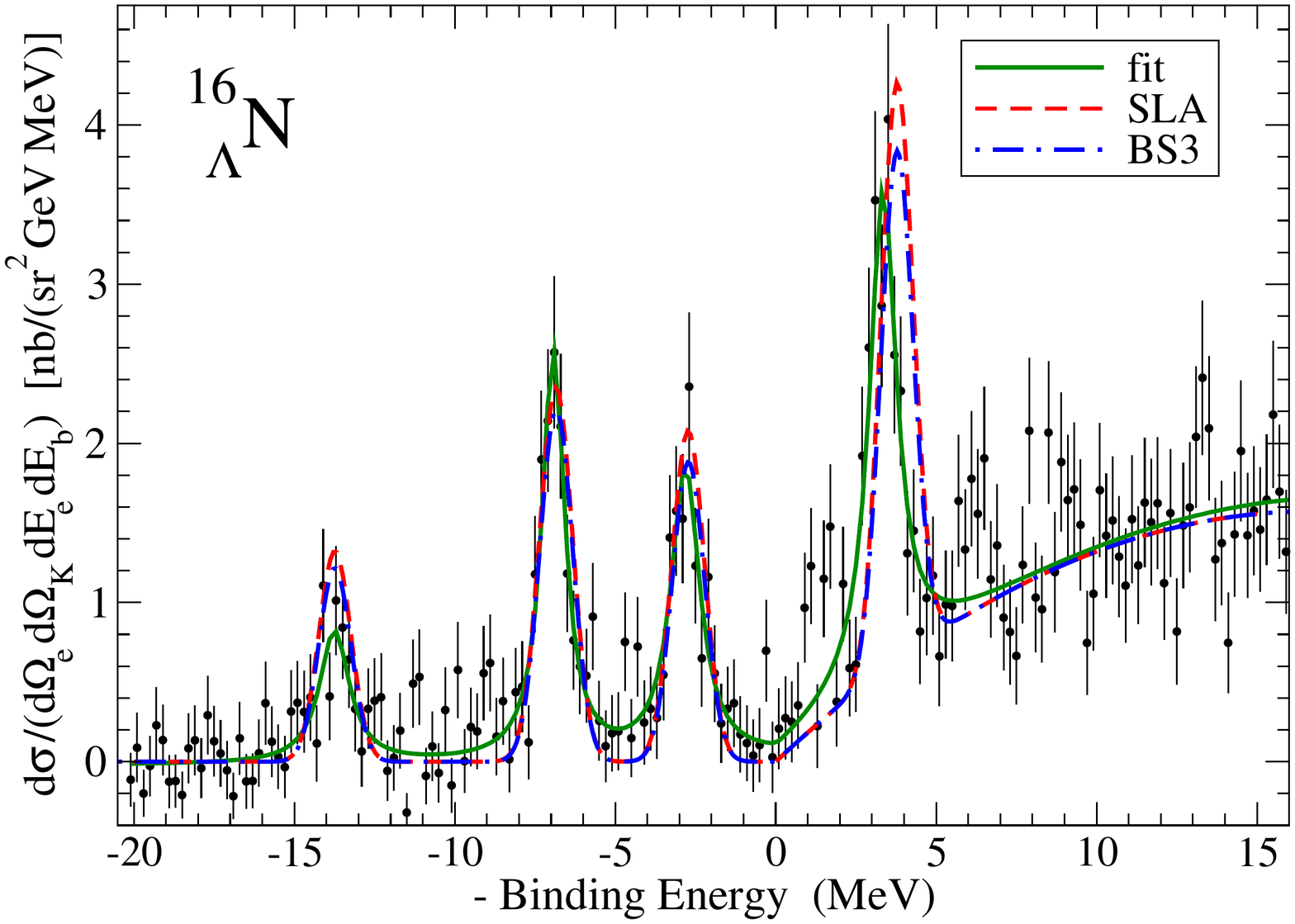}
\caption{$^{16}$O($e,e'K^+$) spectrum of $^{16}_{~\Lambda}$N from JLab Hall A
measurements~~\cite{JLAB09}. Figure adapted from Ref.~\cite{JLAB19}.}
\label{fig:L16N}
\end{figure}

In our previous work we fitted $b_0$ and $B_0$ to the $1s_{\Lambda}$ and 
$1p_{\Lambda}$ states in {\it one} of the nuclear $1p$-shell hypernuclei 
where the $1s_{\Lambda}$ state is bound by over 10~MeV, while the 
$1p_{\Lambda}$ state has just become bound. This helps resolve the density 
dependence of $V_{\Lambda}^{\rm opt}$ by setting a good balance between its 
two components, $V_{\Lambda}^{(2)}(\rho)$ and $V_{\Lambda}^{(3)}(\rho)$, 
following it throughout the periodic table up to the heaviest hypernucleus 
of $^{208}_{~~\Lambda}$Pb produced todate. Among the $A=12,13,16$ relevant 
$1p$-shell hypernuclei, we chose to fit the $^{16}_{~\Lambda}$N precise 
$B^{\rm exp}_{\Lambda}(1s,1p)$ values derived from the first and third peaks, 
respectively, from the left in Fig.~\ref{fig:L16N}. The extremely simple 
$1p$ proton hole structure of the $^{15}$N nuclear core in this case removes 
most of the uncertainty arising from spin-dependent residual $\Lambda N$ 
interactions \cite{Mill08}. 

\section{Results and Discussion}
\label{sec:results}

\subsection{Fits to data}
\label{sub:fits}

In Ref.~\cite{FGa23} we chose to fit only the precise binding energies of 
the $1s_{\Lambda}$ and $1p_{\Lambda}$ states in $^{16}_{~\Lambda}$N and then 
compare predictions with experiment up to $^{208}_{~~\Lambda}$Pb. In the 
present work we performed conventional least-squares fits to the whole data 
of Table~\ref{tab:data} or parts thereof, by varying the two parameters 
of the potential $b_0$ and $B_0$, Eqs.~(\ref{eq:V2}) and~(\ref{eq:V3}) 
respectively. The suppression factor $F$ of Eq.~(\ref{eq:F}) was applied 
when appropriate, namely, for $^{51}_{~\Lambda}$V, $^{89}_{~\Lambda}$Y, 
$^{139}_{~~\Lambda}$La, and $^{208}_{~~\Lambda}$Pb where it varies between
0.85 and 0.67. Whenever $^{12}_{~\Lambda}$B was included in the fits, 
the exceedingly small quoted errors for this hypernucleus as compared with 
the rest of Table~\ref{tab:data} were increased by an order of magnitude 
in order not to distort the results. 

By the nature of the optical potential employed here, namely a sum
of two terms, correlations between the two could be expected. Indeed 
the off diagonal elements of the 2x2 error matrix showed that the two
variables $b_0$ and $B_0$ are 100\% correlated and the error elipse 
then degenerates to a straight line \cite{Lyons86}.

\begin{figure}[!ht]
\begin{center}
\includegraphics[height=8.0cm,width=0.70\textwidth]{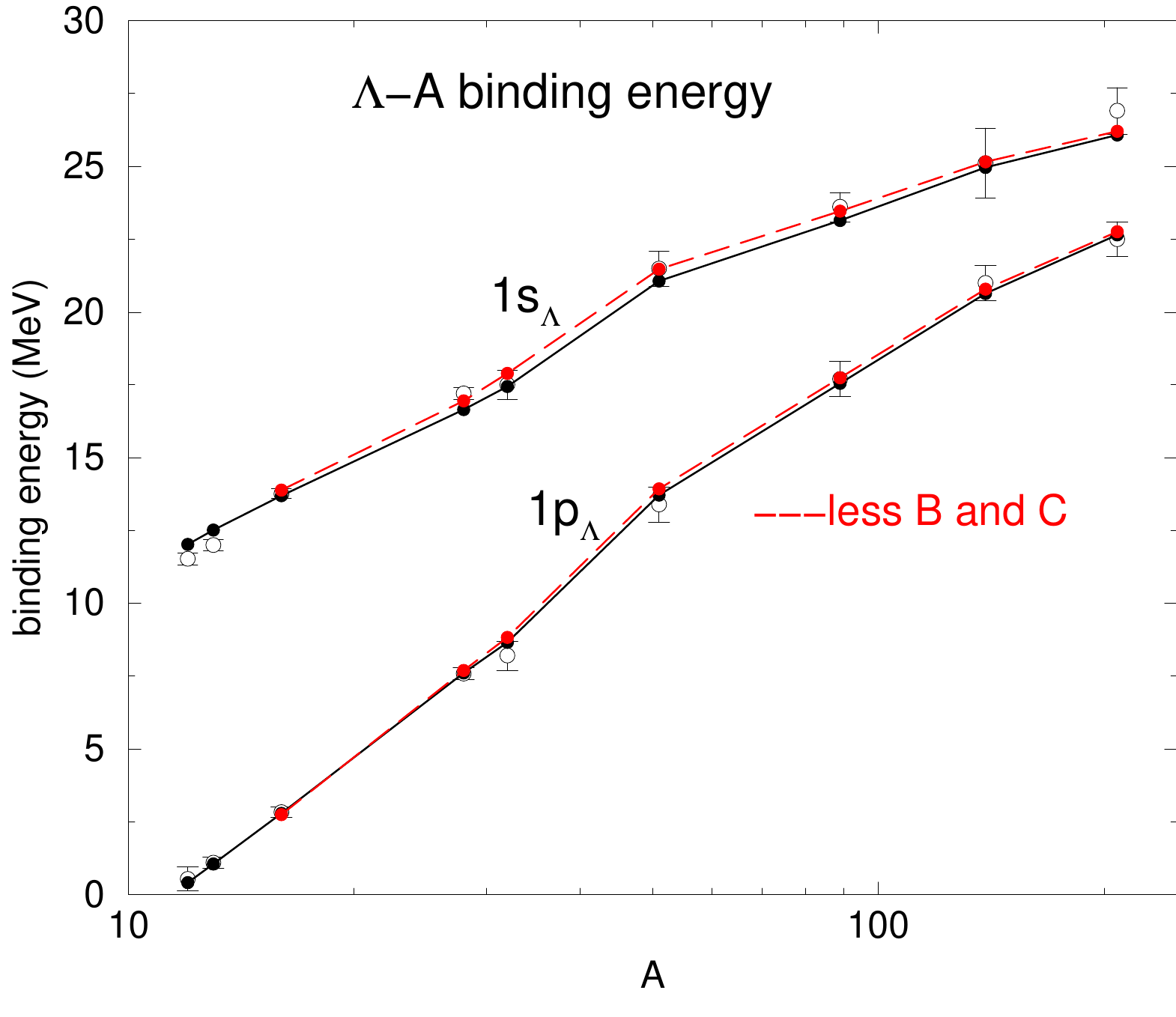}
\caption{Least-squares fits to $B_{\Lambda}$ data using Eqs.~(\ref{eq:V2}) 
and~(\ref{eq:V3}). Black solid lines correspond to the full $B_{\Lambda}$ set 
listed in Table~\ref{tab:data}, red dashed lines correspond to excluding 
$^{12}_{~\Lambda}$B and $^{13}_{~\Lambda}$C. Open circles with error bars 
mark experimental $B_{\Lambda}$ values listed in Table~\ref{tab:data}.}
\label{fig:finalI}
\end{center}
\end{figure}

Figure~\ref{fig:finalI} shows several fits to the $B_{\Lambda}$ data. 
Black solid lines show fits to the full data set of Table~\ref{tab:data}, 
where open circles with error bars mark $B_{\Lambda}$ data points. It is 
clearly seen that the $1s_{\Lambda}$ states in $^{12}_{~\Lambda}$B and 
$^{13}_{~\Lambda}$C do not fit into the otherwise good agreement with 
experiment for the heavier species. The red dashed lines show a very good fit 
obtained upon excluding these light elements from the $B_{\Lambda}$ data set. 
Recall that the experimental uncertainties listed in Table~\ref{tab:data} are 
used unchanged for the red points fit. 

\begin{figure}[!ht]
\begin{center}
\includegraphics[height=8.0cm,width=0.70\textwidth]{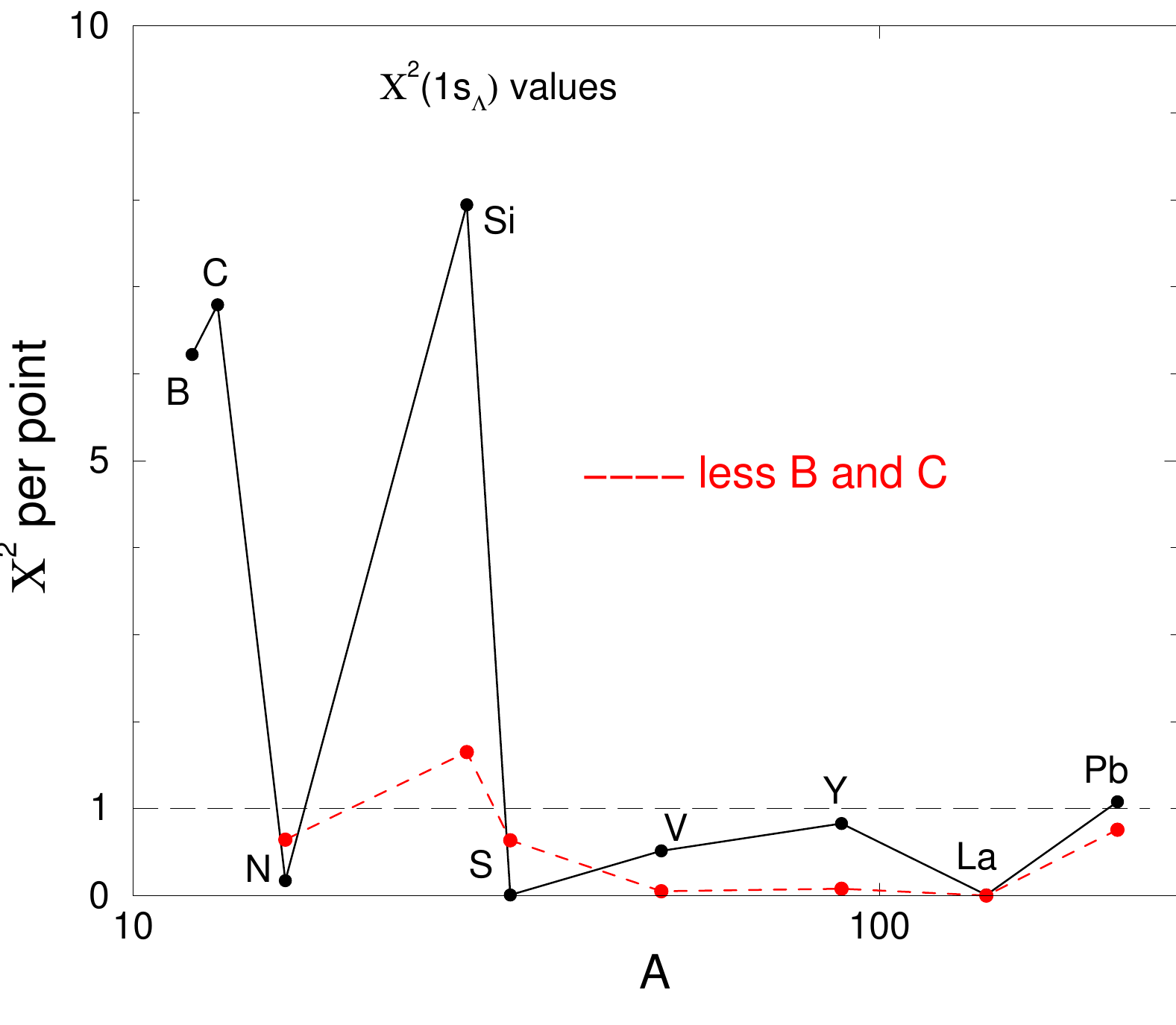}
\caption{Best-fit $\chi ^2$ values for $1s_{\Lambda}$ states. Black solid 
lines correspond to the full data set, red dashed lines correspond to 
excluding $^{12}_{~\Lambda}$B and $^{13}_{~\Lambda}$C, see text.}
\label{fig:finalIX2}
\end{center}
\end{figure}

It is instructive to examine best-fit values of $\chi ^2$ for individual bound 
states. Figure~\ref{fig:finalIX2} depicts such values for the fits shown in
Fig.~\ref{fig:finalI} and it is self evident that by excluding the
binding energies of $^{12}_{~\Lambda}$B
and $^{13}_{~\Lambda}$C the optical potential describes very well the
experimental data. For the $1p_{\Lambda}$ states (not shown here)
the effect of excluding the two lighter species is less pronounced. 

\begin{figure}[!ht]
\begin{center}
\includegraphics[height=8.0cm,width=0.70\textwidth]{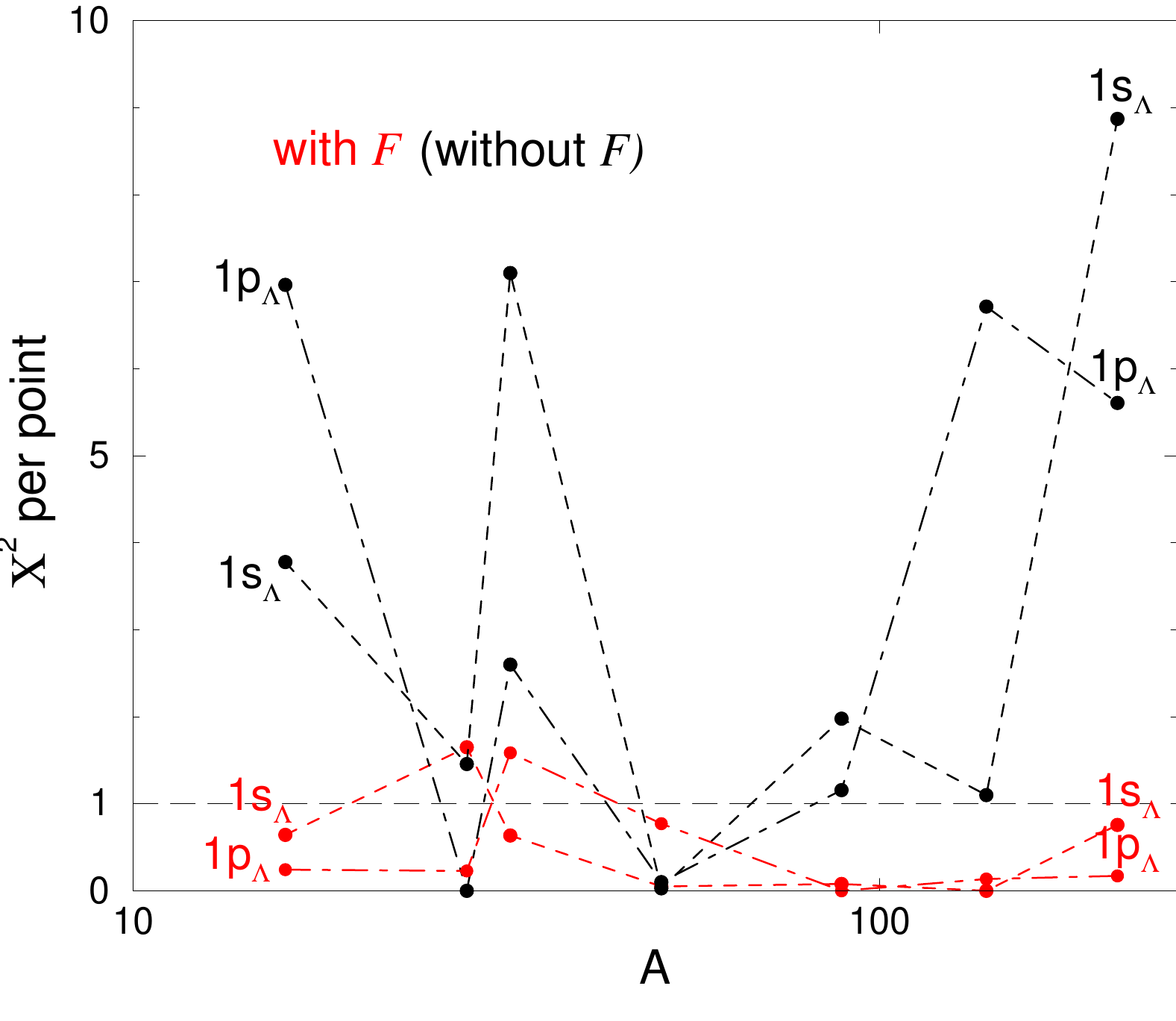}
\caption{Best-fit $\chi ^2$ values for various $1s_\Lambda$ and $1p_\Lambda$ 
states, excluding $^{12}_{~\Lambda}$B and $^{13}_{~\Lambda}$C.
Black lines without the suppression factor $F$ of Eq.~(\ref{eq:F}),
red lines with the $F$ factor, see text.}
\label{fig:X2noX}
\end{center}
\end{figure}

Figure~\ref{fig:X2noX} demonstrates the importance of the suppression 
factor $F$ of Eq.~(\ref{eq:F}) applied to the $\rho ^2$ term of the
potential for medium weight and heavy hypernuclei. Note that the
introduction of this factor does not involve any additional parameter
beyond $b_0$ and $B_0$ and its explicit form, Eq.~(\ref{eq:F}), is based
on a simple shell-model picture. Indeed it was noted in Ref.~\cite{FGa23}, 
and reproduced in Fig.~\ref{fig:FGa23} here, that extrapolating to heavier 
hypernuclei a potential that fits $^{16}_{~\Lambda}$N leads to underbinding, 
unless the suppression factor $F$ is included. 
For the `less B and C' fits of Figs. \ref{fig:finalI} and \ref{fig:finalIX2}
the parameters of the potential 
Eqs.~(\ref{eq:V2}) and (\ref{eq:V3}) are:
\begin{equation}
b_0 = 1.437\,\pm\,0.095~{\rm fm,}\,\,\,\,\,\,\ ({\rm attraction}),
\label{eq:b0}
\end{equation}
\begin{equation}
B_0 = 0.190\,\pm\,0.024~{\rm fm,}\,\,\,\,\,\,\ ({\rm repulsion}).
\label{eq:B0}
\end{equation}
A 100\% correlation between the two parameters holds for the corresponding 
partial potential depths (in MeV):
\begin{equation}
D^{(2)}_\Lambda=-38.6\pm0.8,\,\,\,D^{(3)}_\Lambda=11.3\pm1.4,\,\,\,
D_\Lambda=-27.3\pm0.6 
\label{eq:total}
\end{equation}
at nuclear-matter density $\rho_0=0.17$~fm$^{-3}$.

\subsection{Additional potential terms}
\label{sub:additional}
As stated in the Introduction, the aim of Ref.~\cite{FGa23} and of the present 
work was to see what can be achieved from a simple optical model potential 
constructed by fits to single-particle bound states of $\Lambda$ hypernuclei. 
With least-squares techniques one can examine other possible terms in the 
$\Lambda$-nucleus potential. Guided by the Pauli correlations correction 
to the term linear in the density Eq.~(\ref{eq:Cpauli}) one might look for 
additional powers of $\rho ^{1/3}$, the next being a $\rho ^{5/3}$ term in the 
potential. Repeating least-squares fits to the data (for $^{16}_{~\Lambda}$N 
and above) there was only insignificant reduction in $\chi ^2$ with the extra 
term being ($-0.4\pm1.7) (\rho /\rho_0)^{5/3}$~MeV. The coefficient $b_0$ of 
the linear part of the potential was unaffected by including the additional 
term, but $B_0$, the coefficient of $\rho ^2$ , was much affected and its 
estimated uncertainty increased when three-parameter fits were made.

\subsection{Higher states}
\label{sub:higher}

\begin{figure}[!ht]
\begin{center}
\includegraphics[height=8.0cm,width=0.70\textwidth]{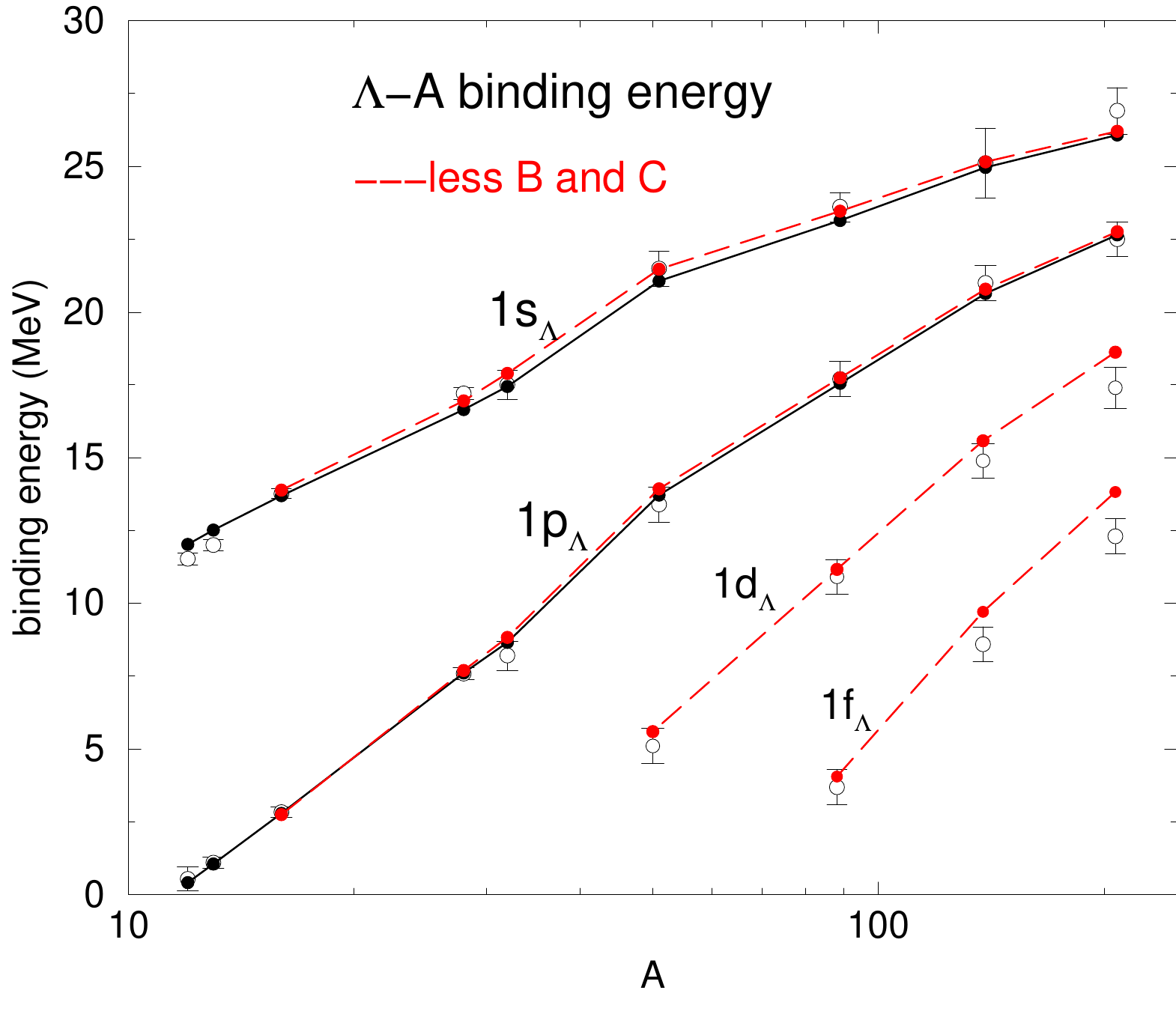} 
\caption{Comparing predictions of binding energies for $1d_\Lambda$ and 
$1f_\Lambda $ states of hypernuclei with experiment. Also shown are best-fit 
calculated energies for $1s_\Lambda $ and $1p_\Lambda $ states, see text.} 
\label{fig:LambdaALL}
\end{center}
\end{figure}

As shown above, very good fits to $1s_{\Lambda}$ and $1p_{\Lambda}$ 
bound states of $\Lambda$ hypernuclei from $^{16}_{~\Lambda}$N to 
$^{208}_{~~\Lambda}$Pb are obtained using a two-parameter optical potential 
of the form given by Eqs.~(\ref{eq:V2}) and (\ref{eq:V3}). Although 
it is not expected that higher states will be well described by the same 
potential, owing to overlooked secondary effects such as non-local terms, 
it is of interest to see to what extent our minimally constructed DD optical 
potential is capable of describing also binding energies of $1d_{\Lambda}$ 
and $1f_{\Lambda}$ states. Experimental values taken from Table~IV of 
Ref.~\cite{GHM16} are displayed in Fig.~\ref{fig:LambdaALL} together with 
predictions made with the best-fit optical potential of the present work. 
It is seen that while slight overbinding of the calculated energies appears 
for the heavier species, the present optical potential reproduces quite well 
the four deepest single-$\Lambda$ bound states in neutron-rich hypernuclei.

\subsection{$1p$-shell nuclei}
\label{sub:1p}

In our preceding work~\cite{FGa23} the two parameters of the potential, $b_0$ 
and $B_0$, were obtained by a fit to the $1s_{\Lambda}$ and $1p_{\Lambda}$ 
binding energies of $^{16}_{~\Lambda}$N. This hypernucleus was chosen 
for two reasons: (i) good precision of the two experimental energies, 
see Table~\ref{tab:data}, and (ii) being at the top end of the $1p$ nuclear 
shell with a single proton hole configuration, only minor effects are 
expected due to other configurations, see Fig.~\ref{fig:L16N}. Indeed, while 
good fits were demonstrated for the heavier species, some difficulties were 
observed when attempting to reproduce experimental $\Lambda$ binding energies 
in the lighter $^{12}_{~\Lambda}$B and $^{13}_{~\Lambda}$C hypernuclei. 
Similar problems were also encountered within the present least-squares 
approach, as demonstrated in Fig.~\ref{fig:finalIX2} above. 
This reflects most likely the insufficiency of reducing a $\Sigma NN$ and 
$\Sigma^{\ast} NN$ mediated $\Lambda NN$ interaction to just one central 
$\rho^2$ term, whereas at least two of the four additional noncentral terms 
are known to be indispensable in $p$-shell $\Lambda$ hypernuclei, mostly in 
the middle of the shell~\cite{GSD71,GSD72,GSD78}.  

Attempts to repeat fits excluding the data for $^{16}_{~\Lambda}$N showed that 
this hypernucleus is quite important in establishing the overall picture 
obtained in the present work. Part of the reason could be the fact that the 
data for the next two species, $^{28}_{~\Lambda}$Si and $^{32}_{~\Lambda}$S, 
are less accurate than the $^{16}_{~\Lambda}$N data and also indicate some 
inconsistency. We note that whereas $^{28}_{~\Lambda}$Si was observed in a 
($\pi^+,K^+$) experiment, $^{32}_{~\Lambda}$S was observed in a ($K^-,\pi^-$) 
experiment, each one undergoing its own energy calibration. 

\begin{table}[!htb]
\begin{center}
\caption{$B_{\Lambda}^{1s}(^{7}_{\Lambda}$Li) in MeV, calculated using 
$V_{\Lambda}^{\rm OPT}$ and a slightly modified $V_{\Lambda}$ form, 
see text, compared to $B_{\Lambda}^{\rm exp}(^{7}_{\Lambda}$Li) from 
emulsion work~\cite{BBF17}.} 
\label{tab:1p} 
\centering 
\begin{tabular}{clll} 
\hline
$_\Lambda^A$Z & ~~~$B_{\Lambda}^{\rm OPT}$ & ~~~$B_{\Lambda}^{\rm modified}$ 
&  ~~~$B_{\Lambda}^{\rm exp}$ \\ 
\hline
$^{7}_{\Lambda}$Li & ~~~5.13$^{+0.25}_{-0.27}$ & ~~~5.30$^{+0.25}_{-0.27}$ & 
~~~5.58$\pm$0.03$\pm$0.04 \\ 
\hline
\end{tabular}
\end{center}
\end{table}

Although the optical-model methodology assumes implicitly a mass number $A$ 
sufficiently large to suppress obvious ${\cal O}(1/A)$ corrections, it is 
instructive to see how the present $V_{\Lambda}^{\rm OPT}$ model fares near 
the beginning of the nuclear $1p$ shell, say in $^{7}_{\Lambda}$Li. Applying 
it with $b_0$ and $B_0$ taken from Eqs.~(\ref{eq:b0}) and (\ref{eq:B0}), 
respectively, gives a $B_{\Lambda}^{1s}$ value listed in the second column of 
Table~\ref{tab:1p}. A straightforward modification of $V_{\Lambda}^{\rm OPT}$ 
would be to remove self-interaction terms in $V_{\Lambda}^{(3)}(\rho)$, 
Eq.~(\ref{eq:V3}), by scaling down $B_0$ according to $B_0 \to (A-1)/A \times 
B_0$ (recall that $A$ stands for the nuclear core mass number) thereby reducing 
the 3-body repulsion term in light hypernuclei. This helps bring the 
$B_{\Lambda}^{\rm modified}$ value listed for $^{7}_{\Lambda}$Li in the 
third column of the table into agreement with the experimental value, 
particularly when recalling the minimal 0.2~MeV uncertainty input adopted 
in our least squares fits.

\subsection{Predictions for upcoming ($e,e'K^+$) experiments on Ca isotopes} 
\label{sub:48K}

$^{40,48}$Ca($e,e'K^+)^{40,48}_{~~~~\Lambda}$K electroproduction 
experiments, aimed at studying single-particle (s.p.) $\Lambda$ spectra 
in $^{40,48}_{~~~~\Lambda}$K, were approved at JLab~\cite{Nakamura22}. 
Calcium targets offer an optimal choice of medium-weight nuclei to explore 
the dependence of the $\Lambda$ s.p. potential on the neutron-excess fraction 
$(N-Z)/A$ which for $^{48}_{~\Lambda}$K is close to 0.2, almost as large as 
in $^{208}$Pb. Calculated $1s_{\Lambda}$ and $1p_{\Lambda}$ binding energies 
in $^{40,48}_{~~~~\Lambda}$K using $V_{\Lambda}^{\rm OPT}$ are listed in 
Table~\ref{tab:K}, where two sets of results appear for $^{48}_{~\Lambda}$K: 
with ($F<1$) and without ($F=1$) applying the neutron-excess suppression 
factor $F$, Eq.~(\ref{eq:F}). Interestingly, the $B_{\Lambda}^{1s}$(OPT) 
values listed for $F=1$ in both $_\Lambda$K isotopes agree very well, up 
to 0.2~MeV, with $B_{\Lambda}^{1s}$(TDA) values calculated very recently 
in Ref.~\cite{Byd23} by applying the Tamm-Dancoff approximation (TDA) to a 
NSC97f-simulated $\Lambda N$ interaction. Yet, the $1p_{\Lambda}-1s_{\Lambda}$ 
excitation energy comes out about 2~MeV higher in TDA than for the present 
$V_{\Lambda}^{\rm OPT}$, questioning thus the applicability of TDA to excited 
$\Lambda$ s.p. states. In contrast, our $B_{\Lambda}(F=1)$ listed values agree 
fairly well, within $\sim$0.3(0.6)~MeV for $1s_{\Lambda}$($1p_{\Lambda}$) 
states, with corresponding values calculated within a quantum Monte Carlo 
method~\cite{LPe17} using considerably stronger $\Lambda N$ and $\Lambda NN$ 
interactions than suggested by our least squares fits. For $F<1$, applying 
the neutron-excess suppression factor, the message of Table~\ref{tab:K} 
is that the resulting $B_{\Lambda}^{1s}$ and $B_{\Lambda}^{1p}$ values are 
$\gtrsim 2$~MeV larger than with no suppression ($F=1$). This message is 
further visualized in Fig.~\ref{fig:dBvsdr2} below. 

\begin{table}[!htb]
\begin{center}
\caption{Calculated values of $B_{\Lambda}^{1s}$ and $B_{\Lambda}^{1p}$ 
in $^{40}_{~\Lambda}$K ($F=1$) and in $^{48}_{~\Lambda}$K ($F=1$ and $F<1$), 
assuming neutron-skin $r_n-r_p$ values of $-$0.04~fm in 
$^{40}_{~\Lambda}$K and 0.16~fm in $^{48}_{~\Lambda}$K, see text.}
\label{tab:K}
\centering
\begin{tabular}{clll}
\hline
$B_\Lambda$~(MeV) & ~~~$^{40}_{~\Lambda}$K~($F=1$) & ~~~$^{48}_{~\Lambda}
$K~($F=1$) & ~~~$^{48}_{~\Lambda}$K~($F=0.69$) \\ 
\hline 
$1s_\Lambda$ & ~~~18.70 & ~~~19.78 & ~~~22.39  \\ 
$1p_\Lambda$ & ~~~10.70 & ~~~12.35 & ~~~14.35  \\ 
\hline
\end{tabular}
\end{center}
\end{table}

\begin{figure}[!ht]
\begin{center}
\includegraphics[height=9.5cm,width=0.80\textwidth]{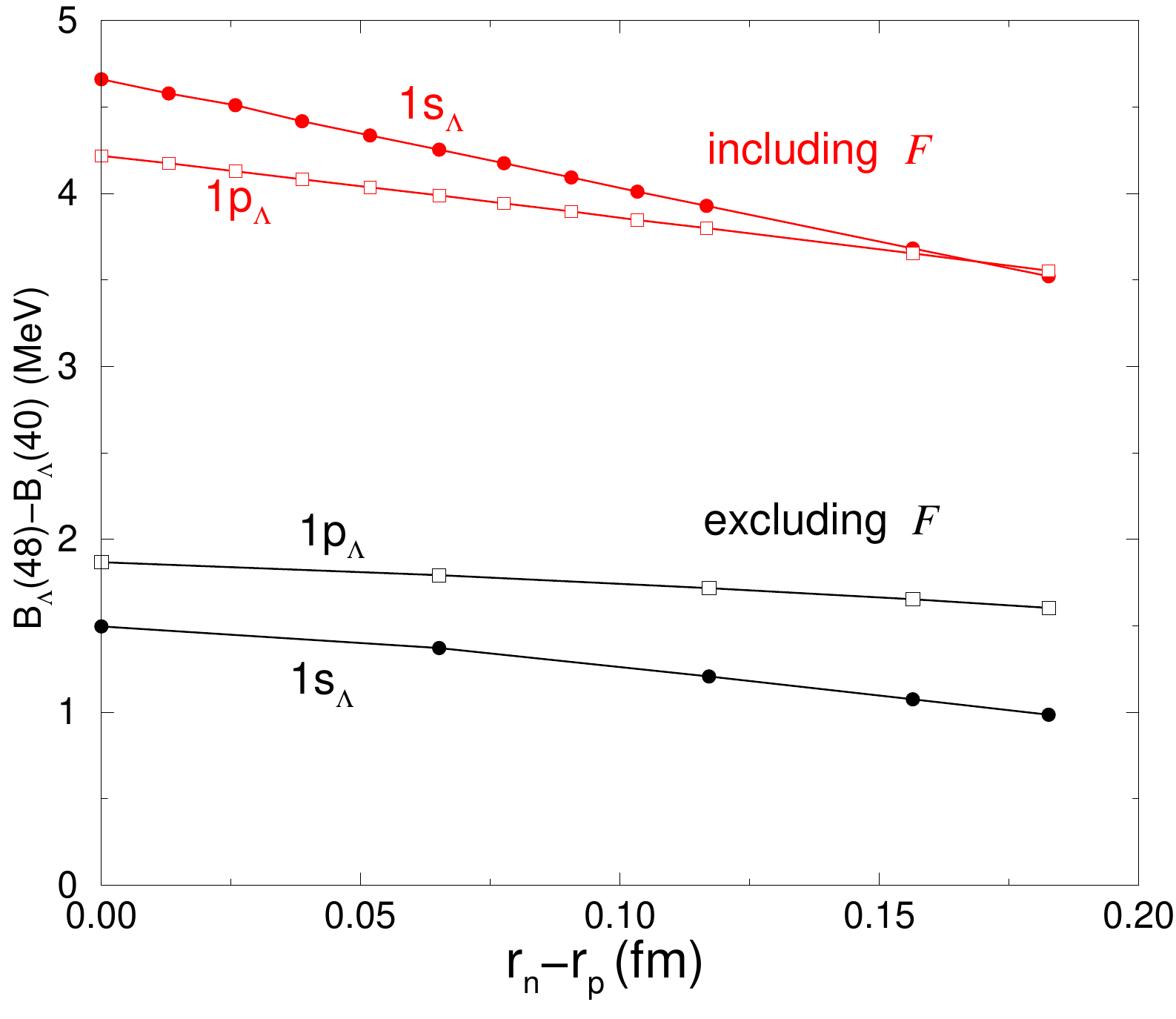}
\caption{Scan over calculated $B_\Lambda$ values in 
$^{48}_{~\Lambda}$K and in $^{40}_{~\Lambda}$K for $1s_{\Lambda}$ and 
$1p_{\Lambda}$ states, with and without applying the suppression factor 
$F$, for neutron densities of $^{48}_{~\Lambda}$K 
characterized by variable neutron-skin $(r_n-r_p)$ values, see text.}
\label{fig:dBvsdr2}
\end{center}
\end{figure}

Figure \ref{fig:dBvsdr2} shows calculated differences of $\Lambda$ binding energies for 
the $1s_{\Lambda}$ and $1p_{\Lambda}$ states between $^{48}_{~\Lambda}$K and 
$^{40}_{~\Lambda}$K as a function of the neutron skin 
$r_n-r_p$, the difference between the r.m.s. radii of the neutron and proton 
distributions in $^{48}_{~\Lambda}$K. 
The figure shows predictions made using our standard $\Lambda$-nucleus 
potential $V_{\Lambda}^{\rm OPT}$ upon including in its upper part (excluding 
in its lower part) the suppression factor $F$, Eq.~(\ref{eq:F}). Regardless of 
the chosen value of $r_n-r_p$, the effect of applying $F$ is about 2.5~MeV for 
the $1s_{\Lambda}$ state and more than 2~MeV for the $1p_{\Lambda}$ state, 
within reach of the upcoming ($e,e'K^+$) approved experiment on $^{40,48}$Ca 
targets at JLab~\cite{Nakamura22}. 

It is worth remarking on the densities used in the above calculations. 
First we note that the charge r.m.s. radii for $^{39}$K and $^{47}$K have 
been measured recently to high precision \cite{KYJ21} and found equal to each 
other within 0.02 fm. We also note that the slopes of {\it all} lines shown in 
Fig.~\ref{fig:dBvsdr2} as a function of $r_n-r_p$ are sufficiently small such 
that an uncertainty of $\pm$0.05~fm in $r_n-r_p$ will cause small uncertainty 
compared to the corresponding shift between the upper and lower sets of 
differences in binding energies. From analyses of typical strong-interaction 
experiments, see e.g. \cite{Fri12} and references therein, we may reliably 
assign a value of 0.15$\pm$0.05 fm for $r_n-r_p$ in $^{48}$Ca, likely to 
provide a good estimate for the corresponding value in $^{48}_{~\Lambda}$K. 
Practically the same value of $r_n-r_p$ in $^{48}$Ca, 0.150$\pm$0.036~fm, 
has been reached in a very recent coupled-cluster calculation~\cite{NLGH23}.

\section{Concluding Remarks} 
\label{sec:disc} 

In the present work we reported on a least-squares fit to $1s_\Lambda$ and 
$1p_\Lambda$ hypernuclear $B_{\Lambda}$ data across the periodic table, $12
\leq A\leq 208$, using a $\Lambda$-nucleus optical potential $V_{\Lambda}^{
\rm OPT}$ specified here in Subsect.~2.1. The optical potential strength 
parameters $b_0$ and $B_0$ of the $\Lambda N$-induced and $\Lambda NN$-induced 
interaction terms, respectively, were fitted to as many as 18 $B_\Lambda$ 
values, allowing unique extraction of the $\Lambda N$ and $\Lambda NN$ 
partial $\Lambda$ potential depths in symmetric nuclear matter at saturation 
density. The values of these potential depths are close to the ones obtained 
by fitting only to $_{~\Lambda}^{16}$N $B_{\Lambda}$ data in our previous 
work~\cite{FGa23,FGa22}. The optical potential strength parameters $b_0$ 
and $B_0$ were found to be 100\% correlated, such that the corresponding 
potential depths $D_{\Lambda}^{(2)}=-38.6\pm 0.8$~MeV and 
$D_{\Lambda}^{(3)}=11.3\pm 1.4$~MeV are also fully correlated, 
with a total potential depth $D_{\Lambda}=-27.3\pm 0.6$~MeV. 

Pauli correlations were found, here too, essential for the balance between 
$b_0$ and $B_0$, as judged by the fitted $b_0=1.44\pm 0.10$~fm getting quite 
close to the value of the $\Lambda N$ spin-averaged $s$-wave scattering length 
(e.g. 1.65~fm \cite{Alex68} or 1.78~fm \cite{Bud10}). Good agreement was 
reached in this model between the $B^{1s,1p}_{\Lambda}$ fit values and 
their corresponding $B^{\rm exp}_{\Lambda}$ values. Although values of 
$\ell_{\Lambda}$ other than $1s_{\Lambda}$ and $1p_{\Lambda}$ were not 
included in these $B_{\Lambda}$ fits, we checked that 
calculated $B_{\Lambda}^{1d,1f}$ 
values in the heaviest available species came out reasonably well. 
Of course, additional, nonlocal (gradient) terms need to be added to 
$V^{\rm OPT}_{\Lambda}$ to achieve better agreement~\cite{MDG88}, but this 
affects little the local terms considered here, and hence the `hyperon puzzle' 
issue. 

Our optical-potential fits to $B_{\Lambda}$ values confirm, 
see Fig.~\ref{fig:X2noX}, the need to suppress $\Lambda NN$ contributions to 
$V^{(3)}_{\Lambda}$ arising from one `excess' neutron and one `core' nucleon. 
Effectively it amounts to replacing $\rho^2(r)$ in $V^{(3)}_{\Lambda}$ by 
$F\,\rho^2(r)$, with a suppression factor $F$ given by Eq.~(\ref{eq:F}). 
This suggests a ${\vec\tau}_1\cdot{\vec\tau}_2$ isospin dependence of the 
underlying $\Lambda N_1N_2$ interaction which follows naturally when $T=1$ 
intermediate hyperons, i.e. $\Sigma$ and $\Sigma^{\ast}(1385)$, connect 
nucleons $N_1$ and $N_2$. We also showed, in the context of upcoming JLab 
($e,e'K^+$) experiments on Ca isotopes, how the suppression implied by our 
$V_{\Lambda}^{\rm OPT}$ affects the $1s_{\Lambda}-1p_{\Lambda}$ spectrum 
in $^{48}_{~\Lambda}$K relative to $^{40}_{~\Lambda}$K. Another JLab approved 
experiment~\cite{Garibaldi22}, $^{208}$Pb($e,e'K^+)\,^{208}_{~~\Lambda}$Tl, 
should test this `suppression' issue, particularly in comparison to a 
$^{208}$Pb($\pi^+,K^+)\,^{208}_{~~\Lambda}$Pb spectrum proposed at the HIHR 
beamline planned at the extended hadron hall of J-PARC~\cite{Nakamura22}. 

The potential depth $D^{(3)}_{\Lambda}$ listed here in Eq.~(\ref{eq:total}) 
agrees within statistical uncertainty with values extracted in our 
previous study~\cite{FGa23,FGa22}. As concluded there, it suggests that the 
$\Lambda$-nucleus potential in symmetric nuclear matter becomes repulsive 
near three times nuclear-matter density $\rho_0$. Our derived depth 
$D^{(3)}_{\Lambda}$ is similar in magnitude to the one yielding 
$\mu(\Lambda) > \mu(n)$ for $\Lambda$ and neutron chemical potentials 
in purely neutron matter under a `decuplet dominance' construction for 
the underlying $\Lambda NN$ interaction terms within a $\chi$EFT(NLO) 
model~\cite{GKW20}. This suggests that the strength of the corresponding
repulsive $V_\Lambda^{(3)}$ optical potential component, as constrained 
in the present work by comprehensive $B_{\Lambda}$ data, is sufficient to 
prevent $\Lambda$ hyperons from playing active role in neutron-star matter, 
thereby enabling a stiff equation of state that supports two solar-mass 
neutron stars.

\section*{Acknowledgments}

We gratefully acknowledge useful remarks by J.~Mare\v{s}, D.J.~Millener,
H.~Tamura, I.~Vida\~{n}a and W.~Weise made on an earlier version presented 
at the HYP2022 International Conference in Prague~\cite{FGa22} as part of 
a project funded by the European Union's Horizon 2020 research \& innovation 
programme, grant agreement 824093.

\end{document}